# An atom interferometer inside a hollow-core photonic crystal fiber

Mingjie Xin, Wui Seng Leong, Zilong Chen, Shau-Yu Lan*

Division of Physics and Applied Physics, School of Physical and Mathematical Sciences, Nanyang Technological University, Singapore 637371, Singapore.

* To whom correspondence should be addressed. Email: sylan@ntu.edu.sg

**Abstract**: Coherent interactions between electromagnetic and matter waves lie at the heart of quantum science and technology. However, the diffraction nature of light has limited the scalability of many atom-light-based quantum systems. We use the optical fields in a hollow-core photonic crystal fiber to spatially split, reflect, and recombine a coherent superposition state of free-falling $^{85}$Rb atoms to realize an inertia-sensitive atom interferometer. The interferometer operates over a diffraction-free distance, and the contrasts and phase shifts at different distances agree within one standard error. The integration of phase coherent photonic and quantum systems here shows great promise to advance the capability of atom interferometers in the field of precision measurement and quantum sensing with miniature design of apparatus and high efficiency of laser power consumption.

# INTRODUCTION

In highly sensitive light-pulse atom interferometers, matter wave beam splitters and mirrors are formed by optical pulses along the atoms' trajectories (*1*). Similar to Ramsey separated oscillatory field method (*2*), the beam splitter and mirror optical pulses serve as local oscillators to compare the phase difference between two interferometer arms via atom-light interaction. In free space, light-pulse atom interferometers are versatile tools for studying fundamental physics (*3-10*) and performing inertial sensing (*11,12*). They have leverage over classical sensors because of the universal properties of atoms and long-term stability (*13*). Nonetheless, the unavoidable bulky and power consuming designs make it very difficult to bring atoms close to surfaces or source masses to maximize signals and thus have limited their use in applications requiring small scale devices or high spatial resolution. For example, to ensure a flat optical wavefront and uniform atom-light interaction, an interferometer beam waist of 2 cm is used in a 10 m-long interferometer which requires 30 W of optical power at 30 GHz detuning from the resonance (*8*). Guiding atoms by light in a micrometer-scale photonic crystal fiber can overcome these obstacles and could enable remote measurement at a targeted location.

Over the past two decades, rapid progress on the design and manufacture of hollow-core photonic crystal fibers provides a nearly single-mode and diffraction-free optical field (*14-16*) for enhancing atom-light interaction such that the confined optical fields can be used for trapping, guiding, and manipulating atoms. Spectroscopic experiments have shown that thermal or cold atoms can be confined or optically guided in hollow-core photonic crystal fibers for quantum and nonlinear optics applications (*17-22*). However, the requirement of the optical wavefront of the fiber for atom interferometer applications is more stringent, where the coherence of optical field in time and spatial domains are crucial for manipulating quantum states of atoms. Here, we optically trap cold $^{85}$Rb atoms in a hollow-core photonic crystal fiber with an $1/e^2$ mode field radius of 22 µm and use the waveguide fields as matter wave beam splitter and mirror pulses to demonstrate that the coherence of a quantum superposition state of atoms can be preserved and interrogated by the optical guided mode to form an interferometer sensitive to the gravity.

**RESULTS**

The experimental setup is shown in Fig. 1. A 4 cm-long hollow-core photonic crystal fiber is mounted vertically inside an ultrahigh vacuum chamber. An ensemble of cold $^{85}$Rb atoms is prepared about 0.5 cm above the fiber tip by a three dimensional magneto-optical trap (MOT). After releasing atoms from the trap at $t\equiv0$, we guide atoms into the fiber by an optical dipole force from the divergent profile of optical field at the facet of the fiber: When the frequency of the light is lower than atomic transition frequency, atoms are attracted toward higher intensity location. With 280 mW of linearly polarized 808 nm dipole trap power, the calculated trap depth is 0.53 mK, assuming 22 µm dipole beam waist. In additiona, we reduce the magnetic field along the vertical axis from 1.90 to 0.61 G to transport atoms closer to the fiber tip to improve the loading efficiency. Push beams with 2 ms duration resonant on both the $F=3$ to $F'=4$ and $F=2$ to $F'=3$ transition are sent horizontally 5 ms before we probe atoms at $t=t_p$ to push atoms above the fiber away from the dipole beam area.

For estimation of atom number inside the fiber, we send a probe beam resonant with the $F=3$ to $F'=4$ transition into the fiber and measure the transmission $T_r$ of the probe beam. Figure 2A shows the optical depth $OD$ versus probe time $t_p$ with and without push beams. The number of atoms inside the fiber can be determined by $OD=-\ln(T_r)=n\sigma L$, where $n$ is the number density of atoms, $\sigma$ is the scattering cross section of atoms, and $L$ is the length of the fiber. Assuming the number density of atoms is Gaussian and integrating over the transverse profile of the fiber, $OD=1$ is approximately equivalent to $1.5\times10^4$ atoms for resonant linearly polarized probe beam (see Materials and Methods). In Figure 2B, we measure *OD* versus the time after releasing atoms from the optical dipole beam and obtain the temperature $T_p$ of the atomic cloud in the radial direction at different probe times from the fitted curves (see Materials and Methods) (*23*). The measurement shows an exponential increase of atom temperature over time which can be explained by heating due to the intensity noise of the dipole beam (*24*).

To demonstrate an atom interferometer, we run a Mach-Zehnder sequence comprising a beam splitter ($\pi/2$) pulse followed by a mirror ($\pi$) pulse and another beam splitter ($\pi/2$) pulse. Atoms are initially prepared in the *F*=2 state by a depumping beam on *F*=3 to *F'*=3 for 1 ms. These mirror and beam splitter pulses are formed by a pair of counter-propagating beams driving Doppler sensitive two-photon Raman transitions between *F*=2 and *F*=3 states. The frequency of the Raman beams is 1.34 GHz red-detuned from the $^{85}$Rb *F*=3 to *F'*=3 transition. With this detuning, we only use about $10^8$ photons for beam splitter pulses. An external magnetic field about 3.6 G is applied along the *z* direction as a quantization axis to break the degeneracy of Zeeman states such that only atoms projected into $m_F$=0 Zeeman state interact with interferometer pulses. After the last pulse, the transition probability of atoms in the *F*=3 state is $(1-\cos(\Delta\varphi))/2$, where $\Delta\varphi=\varphi_1(t_0)-2\varphi_2(t_0+T)+\varphi_3(t_0+2T)$, where $\varphi_i$(i=1,2,3) are the phases of the Raman fields on these three pulses, $t_0$ is the time at the start of the first $\pi/2$ pulse, and *T* is the time separation between interferometer pulses. We only detect atoms in state *F*=3 by the probe beam. When atoms are under free fall, the phases of three pulses are $\varphi_1(t_0)=\omega t_0+\varphi_1^0$, $\varphi_2(t_0+T)=\omega(t_0+T)-(\boldsymbol{\beta_1}-\boldsymbol{\beta_2})\cdot\boldsymbol{g}T^2/2+\varphi_2^0$, and $\varphi_3(t_0+2T)=\omega(t_0+2T)-(\boldsymbol{\beta_1}-\boldsymbol{\beta_2})\cdot\boldsymbol{g}(2T)^2/2+\varphi_3^0$, where $\omega$ is the laser angular frequency, $|\boldsymbol{\beta_1}-\boldsymbol{\beta_2}|\approx 2\beta$ are the propagation constants of the Raman fields in the waveguide, $\boldsymbol{g}$ is the local gravity, and

$\varphi_i^0$ (i=1,2,3) are the initial phases of the Raman fields. Therefore, the phase after closing the interferometer can be written as

$$\Delta\varphi = -(\boldsymbol{\beta_1}-\boldsymbol{\beta_2})\cdot \boldsymbol{g} T^2 + \varphi_1^0 - 2\varphi_2^0 + \varphi_3^0, \qquad (1)$$

where the first term is due to the Doppler shift of the free-falling atoms in the interferometer fields.

Figure 3A shows the interference fringes by varying the laser phase $\varphi_3^0$ of the third pulse at different pulse separation times $T$, whereas the optical dipole trap is off during the interferometer time. The phase shift due to the Doppler shift of free-falling atoms becomes observable at longer $T$. The data are fitted by a sinusoidal function and the contrast is defined by the ratio of the difference and the sum of the maximum and minimum of fitted curves. In Fig. 3B, we extract the phase shifts at different interferometer times by fitting the fringes to a sinusoidal function. We compare the values with our estimation of Eq. 1. The local gravity in Singapore is obtained from International Gravity Formula 1980 (IFG80) as $g$=9.78 m s$^{-2}$ and the propagation constant $\beta=2\pi/(780\times10^{-9})$ m$^{-1}$ is approximated as the wave vector in free space (*25*). The parabolic increase of the phase shift over time indicates that the wavefront inside the hollow-core photonic crystal fiber can be used as a phase reference for coherent quantum system. We plot the contrast versus time $T$ with dipole beam on and off during the interferometer time in Fig. 3C. The decay of contrast with the dipole beam off is mainly due to the ballistic expansion of atoms in the fiber. When the dipole is on during the interferometer time, severe decay of contrast is observed. This is mainly due to the inhomogeneous broadening of differential ac stark shift $2\pi\times3$ kHz between atomic states from the optical dipole beam.

Figure 4A shows interferometer fringes with $T$=100 µs at different probe times. The contrasts and phase shifts agree within one standard error. Using kinematic equation $gt_p^2/2$ from 25 to 35 ms, we operate the interferometer over 3 mm distance inside the fiber. This could be compared to the free space parameter: The Rayleigh length $z_R=\pi W^2/\lambda$ of a Gaussian beam with waist $W=22$ μm is 1.95 mm, where $\lambda$ is the wavelength of the optical field. We also introduce timing asymmetry $\delta T$ between three pulses to measure the velocity width of the atomic ensemble participating in the interferometric process as shown in Fig. 4B (*26*). We plot the contrast $\chi$ versus the offset time $\delta T$ and fit the contrast to a function $\chi(\delta T)=A\exp(-(v_0(\beta_1+\beta_2))^2(\delta T)^2/2)$, where $A$ is the contrast at $\delta T=0$. By assigning $\beta_1 \approx \beta_2 = 2\pi/(780\times10^{-9})$ m$^{-1}$, the fitted curve shows $v_0$=3.17(0.25) cm s$^{-1}$, where $v_0$ is defined as the root mean square velocity of the atomic ensemble in the axial direction. Defining the thermodynamic temperature $T_p=mv_0^2/k_B$, we estimate that the axial temperature of our atoms for the interferometer experiment is about 10 μK, where $m$ is the mass of $^{85}$Rb and $k_B$ is the Boltzmann constant.

## DISCUSSION

Atom numbers and the contrast of our interferometer can be improved by lowering the temperature of the atomic ensemble with Raman sideband cooling and transferring all atoms to $m_F$=0 Zeeman state. The heating due to the dipole intensity noise can be minimized by implementing an intensity stabilization feedback system. To further extend the coherence time, operating our interferometer pulses with Bragg diffraction could eliminate the inhomogeneous broadening of differential ac stark shift from the optical dipole potential on two interferometer arms (*27*). With an improvement of the coherence time to milliseconds, a vibration isolation platform would be required to damp the mechanical noise as in free space interferometers. Our experimental realization of an atom interferometer in a hollow-core photonic crystal fiber could be used to study the mode structure of the fiber which is crucial for understanding decoherence mechanism of the interferometer and also the loss of atom numbers. With 10$^4$ atoms and $T$=1 ms coherence time, our interferometer can have a relative sensitivity of the propagation constant $\Delta\beta/\beta=6\times10^{-5}$ with a distance of the separation of atomic wave packets $2v_rT$~10 μm averaging over the size of the atomic ensemble, where $v_r$ is the photon recoil velocity of the laser used here.

The combination of the photonic system, low power and small volume of our interferometer shows prospects of compactness and versatility. For example, in Newton's constant $G$ measurement (*6*) and the proposed gravitational Aharonov-Bohm experiment (*28*), our interferometer could permit a short distance arrangement between source masses and atoms to enhance the signal. In the study of Rosi *et al.* (*6*), with the same amount of source masses and interferometer performance, our interferometer could increase the phase shift by a factor of 7. More importantly, while the size and the position of the atomic cloud are the leading systematic effects at the level of hundreds of micrometers, our interferometer can have an improvement by two orders of magnitude with 10% uncertainty of the fiber position.

The precision (or uncertainty) of any atom interferometers is fundamentally limited by the atom shot noise. The precision of Rosi *et al.* (*6*) is 0.012 rad. with $2\times10^5$ atoms used, a factor of 5 away from atom shot noise. Although our fiber atom interferometer only uses about $10^4$ atoms, orders of magnitude improvement can be made in the future by cooling atoms to lower temperature and optical pumping to the magnetic field-insensitive state. By preparing optically thick atomic medium in the fiber, spin-squeezing techniques can be applied to improve the sensitivity of the interferometer (*29,30*). Our experiment demonstrated here will also initiate a campaign of studying the systematic effects of atom interferometers in the hollow-core fiber such as the propagation constant and imperfection of the fiber mode structure. In conclusion, our realization of an atom interferometer inside a hollow-core photonic crystal fiber indicates coherent interaction of the optical waveguide mode and quantum system, and could open up novel applications for precision measurement and quantum sensing.

# MATERIALS AND METHODS

## Atom loading

Atoms released from a dispenser are trapped and cooled by a three-dimensional MOT. After 580 ms, the external magnetic field along the *z* axis is reduced from 1.90 to about 0.61 G to bring the atomic ensemble closer to the fiber tip. The magnetic fields along the *x* and *y* axes are adjusted to position the atomic cloud to maximize the number of atoms loaded into the fiber. When all the optical fields for cooling are off, the optical dipole potential and the gravity take over to attract atoms into the fiber. The push beams are on for 2 ms right before the probe beam. The fiber used in the experiment is from GLOPhotonics (PMC-C-TiSa_Er-7C). It is a hypocycloid-shaped photonic crystal fiber. The $1/e^2$ mode field radius $W$ is 22 μm. The optical dipole potential is provided by a Ti-Sa laser at 808 nm. The trapping potential at the center of the dipole trap inside the fiber is about 0.53 mK for our measurement.

## Detection method

To measure *OD*, we sent two 3 nW probe pulses with 50 μs duration, one for detecting atoms and the other one for reference. The second pulse is sent 30 ms after the first pulse to ensure that atoms diffuse out of the probe beam area. The probe power is chosen to avoid saturating the atoms because the saturation power, averaging over the transitions, is ~30 nW. To avoid any transient effects of the pulses, we selected only the middle 40 μs of the pulses for *OD* measurement. The probe beam is linearly polarized and 8.3 MHz blue-detuned from $F=3$ to $F'=4$ for data in Fig. 2. It is then detected by an avalanche photodiode from Hamamatsu (C10508-01). We estimated the number of atoms loaded into the fiber by integrating

$$OD = (2/(\pi W^2))L\int_0^R n_0 \exp(-r^2/r_0^2)\sigma\exp(-2r^2/W^2)2\pi r dr \qquad (2)$$

$$= (2\sigma n_0 L\, r_0^2/(W^2+2r_0^2))(1-\exp(-R^2(W^2+2r_0^2)/(Wr_0)^2)),$$

where $\sigma=\sigma_0/(1+4(\delta_p/\Gamma)^2)$ is the scattering cross section of atoms at the center of the dipole trap, $\sigma_0$ is the resonant cross section, $\delta_p$ is the detuning of the probe beam, $\Gamma$=6.1 MHz is the decay rate of the excited state, $r_0$ is the 1/$e$ radius of the initial atomic cloud size, $r$ is the radial coordinate, $R$ is the radius of the fiber core, and $n_0$ is the peak atomic density. The radius of the atomic cloud can be inferred from the ratio of the cloud temperature $T_p$ and trapping potential $U_0$ as $r_0^2=(k_B T_p/(2U_0))W^2$. For $T_p$=23μK, $U_0/k_B$=0.53 mK, linearly polarized probe beam resonant on $m$=0 to $m'$=0 Zeeman transition, and $\sigma_0$=0.53×10$^{-9}$ cm$^2$, arrived at by averaging over all allowed transitions between $F$=3 and $F'$=4 Zeeman states, $OD$=1 corresponds to about 1.5×10$^4$ atoms.

**Temperature measurement in the radial direction**
When atoms are inside the fiber, the optical dipole trap is switched off, allowing the atomic cloud to expand freely. After a release time, we sent the probe beams for $OD$ measurement as shown in Fig. 2B. We considered the ballistic expansion of the atomic cloud and thus replaced $r_0^2$ by $r_0^2(t)=r_0^2+v_p^2 t^2$ in Eq. 2. To reduce the parameters for fitting, we further used $v_p^2=2k_B T_p/m$ and $r_0^2=W^2 k_B T_p/2/U_0$ in Eq. 2. With $U_0$=0.53 mK, $W$=22 μm, and $R$=31.5 μm used in the experiment, we fit the data with an equation

$$OD(t)=OD_0(1-\exp(-C(1/(AT_p+BT_p t^2)+1)))(1/(AT_p+BT_p t^2+1)), \qquad (3)$$

where $OD_0$ is the optical depth when the dipole trap is turned off, $A= k_B/U_0$=0.0019 μK$^{-1}$, $B= 4k_B/m/W^2$=0.79 μK$^{-1}$ms$^{-2}$, and $C=2R^2/W^2$=4.1. The fitting results are shown in Fig. 2B.

**Interferometer**

Our interferometer pulses are generated from a diode laser locked to $F=2$ to $F'=3$ of $^{87}$Rb which is 1.34 GHz red-detuned from the $^{85}$Rb $F=3$ to $F'=3$ transition. The Raman fields are formed by a fiber EOM modulated around $^{85}$Rb ground state hyperfine splitting frequency. The 0-order and the +1-order of the fields after EOM are used for the two-photon Raman transition. Although 0-order and -1-order also form another pair of two-photon transition with opposite phase, its larger single photon detuning will give a reduction factor of $\omega_{hf}/(1.34\text{ GHz}+\omega_{hf})\approx 0.69$ on the effective two photon Rabi frequency (*31*), where $\omega_{hf} = 3.036$ GHz is the energy splitting of hyperfine ground states of $^{85}$Rb. The fields after passing through EOM are split into two optical paths and coupled into the hollow-core fiber from both ends. We further tuned this frequency to match the Doppler shift of atoms. The polarization of the counter-propagating fields is linear and orthogonal to each other. The optical power in the top and bottom Raman beams is 4 µW and 8 µW, respectively. We shifted the phase of the microwave sent to the EOM during the last interferometer pulse. The probe beam in the interferometric measurement is resonant on $F=3$ to $F'=4$.

**Acknowledgments:** We thank Pei-Chen Kuan, Holger Müller, and Sheng-wey Chiow for stimulating discussions and reading the manuscript. We thank Arpan Roy and Wei Sheng Chan for their contribution to the apparatus. **Funding:** This work was supported by the Singapore National Research Foundation under grant no. NRFF2013-12 and the Nanyang Technological University under start-up grants. **Author contributions:** All authors contributed to designing the experiment, analyzing the data, and preparing the manuscript. **Competing interests:** The authors declare that they have no competing interests. **Data and materials availability:** All data needed to evaluate the conclusions in the paper are present in the paper. Additional data related to this paper may be requested from the authors.

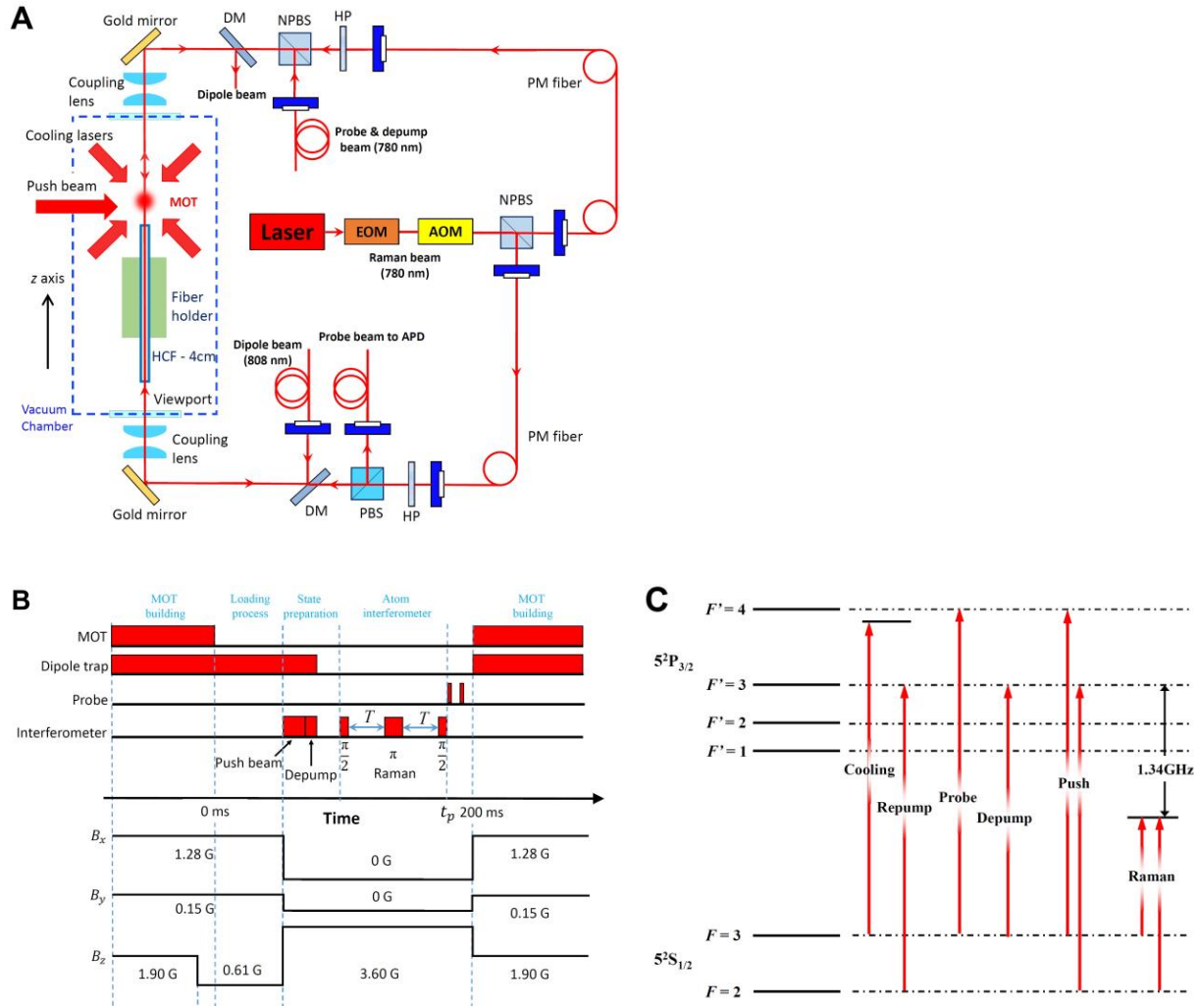

**Fig. 1. Experimental configuration.** (**A**) Experimental setup. NPBS: nonpolarizing beam splitter. PBS: polarizing beam splitter. DM: dichroic mirror. HP: half-wave plate. PM fiber: polarization maintaining fiber. EOM: electro-optical modulator. AOM: acousto-optical modulator. APD: avalanche photodiode. HCF: hollow-core fiber. (**B**) Timing sequence for atom interferometer (not drawn to scale). Each experimental cycle takes 800 ms, including 580 ms for cooling atoms. The external magnetic fields $B_x$, $B_y$, and $B_z$ are for canceling the ambient magnetic field, shifting the atomic ensemble, and defining quantization axis for atom-light interaction. (**C**) Relevant energy level diagram showing frequencies of laser beams used.

A

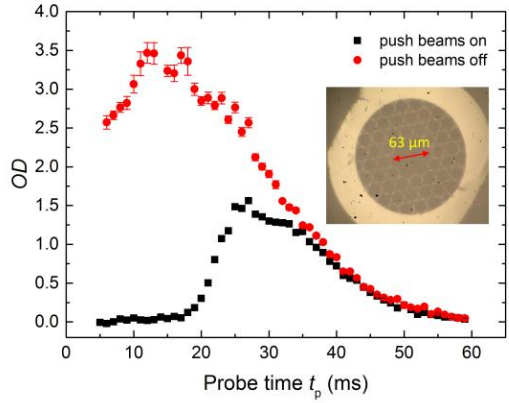

B

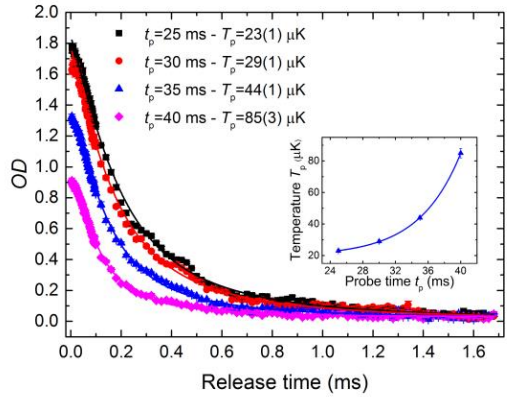

**Fig. 2. Atoms inside the fiber.** (**A**) *OD* versus probe time with push beams on and off. The probe time at $t_p$=0 corresponds to the time of releasing atoms from the MOT. Atoms are reloaded for each data point and the optical dipole trap is off when the probe field is on. The error bars represent the standard error of five experimental runs. The inset shows an image of the cross-section of the hollow-core fiber used in the experiment. (**B**) Measurement of atom temperature $T_p$ in the radial direction at different probe times (see Materials and Methods). The release time in the *x* axis is the duration where the dipole trap is off for atoms to expand. The error bars indicate the standard error of seven experimental runs. The inset is a plot of temperature at different probe times.

**A**

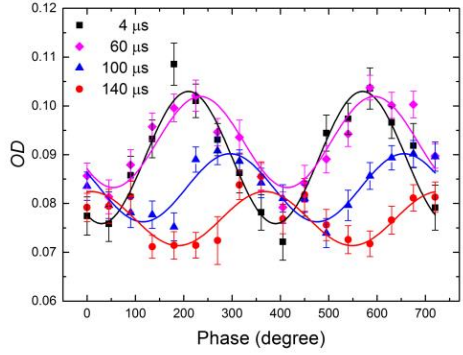

**B**

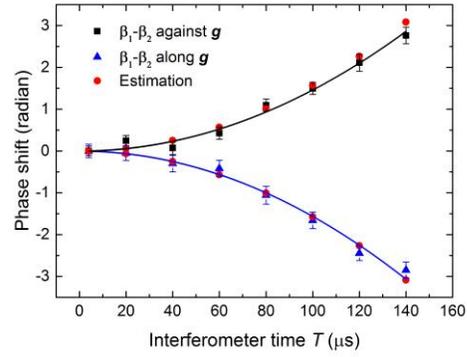

**C**

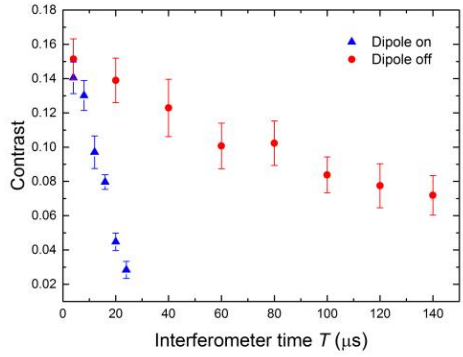

**Fig. 3. Mach-Zehnder interferometer.** (**A**) *OD* versus phase with different interferometer times *T*. The fringes are fitted to a sinusoidal function. Each data point is an average of 60 runs with the standard error. (**B**) Phase shift versus time *T*. Each data point is extracted from the fitting of a sinusoidal function as in (A) with the Raman beam direction along and opposite to the gravity. The zero point in the *y* axis corresponds to the phase at *T*=4 μs. The curves are the fits with Eq. 1. The error bars are the uncertainty of the sinusoidal fits. The red circles are the phase shifts based on the estimation of Eq. 1. (**C**) Contrast versus time. A comparison of the decay time with optical dipole trap on until $t_p$ and off during the interferometer sequence. The error bars are from the uncertainty of the sinusoidal fits.

**A**

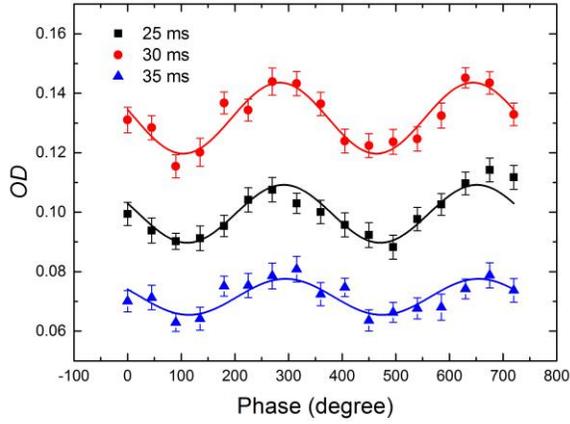

**B**

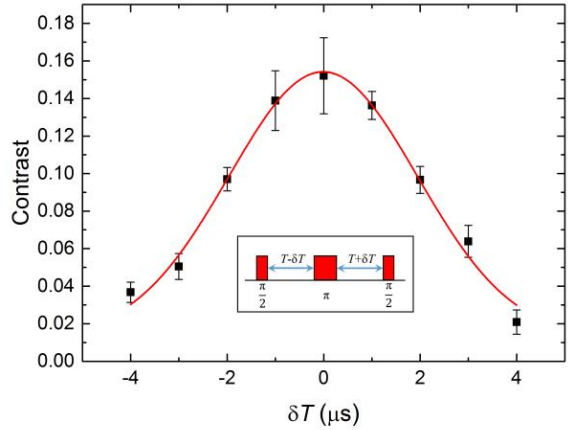

**Fig. 4. Interferometer at different locations of the fiber.** (**A**) *OD* versus phase at different probe times. Data are fitted to a sinusoidal function. The standard error is an average of 60 experimental runs. (**B**) Velocity width measurement of atoms selected in the interferometer measurement. We use the sequence as shown in the inset. Each data point is from a sinusoidal fitting of interference fringe similar to Fig. 3A with $T$=4 μs. The contrast decay is fitted to a Gaussian function. The error bars are from the uncertainty of the sinusoidal fitting.